\documentclass[sigconf]{acmart}
\AtBeginDocument{%
  \providecommand\BibTeX{{%
    \normalfont B\kern-0.5em{\scshape i\kern-0.25em b}\kern-0.8em\TeX}}}

\copyrightyear{2022}
\acmYear{2022}
\setcopyright{acmlicensed}\acmConference[FAccT '22]{2022 ACM Conference on Fairness, Accountability, and Transparency}{June 21--24, 2022}{Seoul, Republic of Korea}
\acmBooktitle{2022 ACM Conference on Fairness, Accountability, and Transparency (FAccT '22), June 21--24, 2022, Seoul, Republic of Korea}
\acmPrice{15.00}
\acmDOI{10.1145/3531146.3533150}
\acmISBN{978-1-4503-9352-2/22/06}
    
\usepackage{enumitem}

\newcommand{\boldedpoint}[1]{\vspace{2mm}\noindent\textbf{#1.}}

\begin{document}

\title{Accountability in an Algorithmic Society:
Relationality, Responsibility, and Robustness in Machine Learning}


\author{A. Feder Cooper}
\authornote{Equal contribution}
\affiliation{%
   \institution{Cornell University}
   \city{Ithaca}
   \state{NY}
   \country{USA}}
   \email{afc78@cornell.edu}

\author{Emanuel Moss}
\authornotemark[1]
\affiliation{%
  \institution{Cornell Tech and Data \& Society Research Institute}
  \city{New York}
  \state{NY}
  \country{USA}
}
\email{emanuelmoss@cornell.edu}

\author{Benjamin Laufer}
\affiliation{%
  \institution{Cornell Tech}
  \city{New York}
  \state{NY}
  \country{USA}}
\email{bdl56@cornell.edu}

\author{Helen Nissenbaum}
\affiliation{%
  \institution{Cornell Tech}
  \city{New York}
  \state{NY}
  \country{USA}}
\email{hn288@cornell.edu}

\renewcommand{\shortauthors}{Cooper and Moss, et al.}
\renewcommand{\shorttitle}{Accountability in an Algorithmic Society}

\begin{abstract}
In 1996, \emph{Accountability in a Computerized Society}~\cite{nissenbaum1996accountability} issued a clarion call concerning the erosion of accountability in society due to the ubiquitous delegation of consequential functions to computerized systems. 
\citet{nissenbaum1996accountability} described four 
barriers to accountability that computerization presented, 
which we revisit in relation to the 
ascendance of data-driven algorithmic systems---i.e., machine learning 
or artificial intelligence
---to uncover 
new challenges for accountability that these systems present. Nissenbaum's original paper grounded discussion of the barriers in moral philosophy; we bring this analysis together with recent scholarship on relational accountability frameworks and discuss how the barriers present difficulties for instantiating 
a unified moral, relational framework in practice for data-driven algorithmic systems. We conclude by discussing ways of weakening the barriers in order to do so. 
\end{abstract}



\maketitle

\section{Introduction}\label{sec:intro}

In 1996, 
\citet{nissenbaum1996accountability}
warned of the erosion of accountability due to four barriers inimical to societies increasingly reliant on computerized systems.
These barriers are: \textit{many hands}, 
to refer to the problem of attributing moral responsibility for outcomes caused by multiple moral actors; ``\textit{bugs},'' the way software developers might shrug off responsibility by suggesting software errors are unavoidable; \textit{computer as scapegoat}, the shifting of blame to computers as if they were moral actors; and \textit{ownership without liability}, the free pass to the software industry to deny responsibility, particularly via shrink-wrap and click-wrap Terms of Service agreements. Today, twenty-five years later, significant work has been done to address the four barriers 
through developments in professional practices of computer science~\cite{vandorp2002tracking, jarke1998requirements}, organizational management~\cite{javed2014systematic}, and civil law~\cite{mulligan2018governance, eu2017civil}; however, 
the effort to restore accountability remains incomplete. In the interim, the nature of computerized systems has 
been radically transformed by the ascendance of data-driven algorithmic systems\footnote{Since rule-based software systems are also ``algorithmic,'' we 
specify which of the meanings we intend in settings where the context 
does not disambiguate.}---e.g., machine learning (ML) and artificial intelligence (AI)---which either have replaced or complemented rule-based software systems, or have been incorporated within them as essential elements~\cite{carbin2019overparameterization, zhang2003machine,barstow1988artificial, mulligan2019ml, kroll2017accountable}.

The resurgent interest in accountability is therefore timely for a world in which data-driven algorithmic systems are ubiquitous.\footnote{We adopt the term ``Algorithmic Society'' as used in~\citet{balkin2018society}.} In domains as varied as finance, criminal justice, medicine, advertising, 
hiring, manufacturing, and agriculture, these systems are simultaneously treated as revolutionary, adopted 
in high-stakes decision software and machines~\cite{angwin2016machine, ajunwa2021auditing, kleinberg2018human, aaj2017avs, moradi2020ai}, and as novelties~\cite{kleeman2016watson}. The failure to comprehensively establish accountability within computational systems through the 1990s and 2000s has  
thus left contemporary societies just as vulnerable to the dissipation of accountability, with even more at stake. We remain in need of conceptual, technical, and institutional mechanisms to assess how to achieve accountability for the harmful consequences of data-driven algorithmic systems---mechanisms that address both \emph{whom} to hold accountable and \emph{how} to hold them accountable for the legally cognizable harms of injury, property loss, and workplace hazards, and the not-yet-legally-cognizable harms increasingly associated with 
these systems, such as privacy violations~\cite{citron2022privacy}, manipulative practices~\cite{agarwal2019deepfake, kreps2020gpt}, and automation-driven 
discrimination~\cite{ajunwa2021auditing}. 

In light of growing concerns over accountability in computing, our paper revisits Nissen\-baum's ``four barriers to accountability'' to assess whether insights from that work remain relevant to data-driven algorithmic systems, and to consider how the ascendance of such systems complicates, challenges, and demands more of sociotechnical, philosophical, and regulatory work. We first provide context on recent developments in standards of care, law and policy, and computer science that are necessary for our analysis (Section~\ref{sec:rw}). Equipped with this background, we recapitulate the elements of moral philosophy on which \citet{nissenbaum1996accountability} depended (Section~\ref{sec:rw:philosophy}), and discuss how this moral conception of accountability can be unified with Bovens's relational definition of accountability in political theory~\cite{bovens2007analysing}, which has drawn recent attention in AI ethics scholarship. In particular, we contend that moral and relational accountability can be brought together to illuminate the necessary parameters of an accountability framework for data-driven algorithmic systems---determining \emph{who} is accountable, \textit{for what}, \textit{to whom}, and \emph{under which circumstances} (Section~\ref{sec:rw:narrow}). To instantiate such a framework, however, requires recognizing the ways in which data-driven algorithmic systems specifically make determining these parameters challenging. We therefore update Nissenbaum's four barriers to accountability in relation to these 
systems, and clarify the ways that each barrier obscures and complicates realizing a moral, relational accountability framework in practice (Section~\ref{sec:barriers}). Finally, we conclude by suggesting ways of weakening these barriers to accountability, thereby strengthening accountability practices for the entire field (Section~\ref{sec:beyond}).

\subsection{Contemporary Interventions in Technological Accountability} 
\label{sec:rw}

Re-visiting the four barriers requires engaging with the significant body of work on accountability produced in the interim. Rather than comprehensively reviewing existing literature---an 
undertaking already addressed in, e.g.,~\citet{wieringa2020account} and~\citet{kohli2018translation}---we highlight three areas of work that we find useful for our analysis
:

\boldedpoint{Standards of care} These 
play a crucial role in building a culture of accountability--- establishing best practices and 
formal guidelines for ensuring 
that concrete practices align with agreed-upon values (e.g., safety).  
In engineering, standards of care dictate the behaviors and outputs expected of sound work. For data-driven algorithmic systems in particular, 
they have taken the form of 
annotations~\cite{beretta2021detecting}, audits~\cite{ajunwa2021auditing}, and frameworks  concerning the appropriate use of data and other artifacts, which are often developed and used in the production of AI/ML systems~\cite{mcmillan2021reusable, boyd2021datasheets, mitchell2019model, shen2021value, hutchinson2021towards, gebru2021datasheets}. Taken together, these standards of care support accountability by making the intentions and expectations around such systems concrete; they provide a baseline against which one can evaluate deviations from expected behavior and, accordingly, are used to review and contest the legitimacy of specific applications of data-driven techniques. 
Some scholars have re-framed 
such standards around harmed and vulnerable parties~\cite{raji2020closing, metcalf2021algorithmic}. This work 
makes clear that standards of care
, while important for developing actionable notions of accountability, do not guarantee accountability on their own~\cite{vecchione2021algorithmic, diakopoulos2020accountability}. Algorithmic impact assessments 
attempt to fill this gap~\cite{moss2021assembling}. They task practitioners with assessing new technologies in terms of their anticipated impacts~\cite{selbst2021institutional, metcalf2021algorithmic}, and 
formalize accountability relationships in ways that may systematically address and correct algorithmic harms.

\boldedpoint{Law and policy} 
Literature on data-driven algorithmic systems generally concerns AI/ML-related harms and corresponding interventions. 
Work on liability spans both anticipated harms related to new or forthcoming data-driven technology, including autonomous vehicles and robotics~\cite{abraham2019responsibility, aaj2017avs, surden2016avs, elish2019crumple, cooper2022avs}, and not-yet-legally-cognizable harms, such as unfair discrimination due to demographically\--imbalanced, biased, or otherwise-discredited training data~\cite{hellman2021data, okidebe2022data, waldman2019power}, privacy violations~\cite{crawford2014harms, citron2022privacy, kaminski2019algorithmic}, and manipulation~\cite{kreps2020gpt}. Regulatory and administrative scholarship tends to analyze data-driven algorithmic systems in relation to legislation and policy that predates many AI/ML technological developments~\cite{shah2018algorithmic, viljoen2021relational, whittington2015push, sadowski2021everyone, mulligan2019ml, delgado2022uncommontask}. That said, recent regulatory interventions, including 
GDPR (the nascent, yet wide-reaching data-privacy policy in the EU~\cite{kang2020algorithmic, wachter2017transparent, hamon2021impossible}) and the California Consumer Privacy Act of 2018~\cite{barrett2019eu}, which have also been applied to AI/ML systems, are increasingly represented within the law and policy literature. 

Law and policy approaches tend to focus on transparency, which is of broad import in democratic governance and is intimately connected to accountability~\cite{mulligan2018governance}. Transparency is necessary for identifying responsible parties (in order to attribute harms to those who are responsible for them), and necessary for identifying the sources of these harms and potential mitigations~\cite{diakopoulos2020accountability}.  
Work in this area spans a range of urgent concerns surrounding lack of transparency in data-driven algorithmic systems. These include the obfuscation of data provenance~\cite{wachter2018bigdata, levy2016transparency}, particularly caused by the concentration of data ownership within data brokers~\cite{ftc2014databrokers,lambdan2019databrokers, young2019beyond}, and insufficient transparency of algorithms and models, which contributes to the inscrutability of automated decisions~\cite{cofone2019secrecy, lehr2017legalml, kroll2017accountable}. Critics have argued that outsourcing legal decisions to automated tools, particularly data-driven tools that obscure underlying decision logic, can create a crisis of legitimacy in democratic decision-making~\cite{mulligan2019ml, gao2021framework, calo2021modeling, citron2020workingpaper}.

\boldedpoint{Computer science} Research 
in AI/ML has increasingly treated accountability as a topic for scholarly inquiry. In updating Nissenbaum's barriers, we address cases in which researchers explicitly recognize the relationship between their work and accountability~\cite{kim2021accountability}---namely, in auditing and transparency---and work on \textit{robustness}, which we identify as having significant implications for accountability, even when this  work itself does not explicitly make the connection. Recent work on audits underscores the importance of being able to analyze algorithmic outputs to detect and correct for the harm of unfair discrimination~\cite{adler2018auditing, raji2020closing}. Transparency tends to be treated as a property of models, particularly whether a model is interpretable or explainable to relevant stakeholders~\cite{doshivelez2018interpretability, freitas2014interpretability,bhatt2021uncertainty}. 
More recently, computational work has begun to take a more expansive view of transparency, applying it to other parts of the ML pipeline, such as problem formulation, data provenance, and model selection choices~\cite{forde2021disparate, kroll2021traceability, sivaprasad2020hpo, sloane2020participation, cooper2021hpo}. 

Lastly, often overlooked, \textit{robustness} 
draws attention to whether a model behaves as expected under 
likely, unlikely, anomalous, or adversarial conditions. Designing for and evaluating robustness implicates accountability, as it requires researchers to define their expectations of model performance rigorously; this in turn encourages inquiry into how 
to prevent deviations from those expectations, and to identify (and ideally correct for) such deviations. Robustness thus encompasses work in AI/ML that aims to achieve 
theoretical guarantees in practice~\cite{yang2019correctness, meinke2019provable, zhang2020tunamh}, and work that, even in the absence of such guarantees, produces models 
with reproducible empirical behavior 
~\cite{bouthillier2019reproducibility, raff2019reproducibility}. Robustness also includes the ability for 
models to generalize beyond the data on which they were trained~\cite{neyshabur2017generalization, hu2020generalization}, ranging from natural cases of distribution shift~\cite{ovadia2019shift, koh2021shift} to handling the presence of adversaries that are trying to  
game model outputs~\cite{goodfellow2014adversarial, papernot2015adversarial,szegedy2013adversarial}.

\section{Conceptual Framing} \label{sec:framing}

The conceptual framing of accountability for this paper draws from two sources of scholarship: 1) moral philosophy, which construes accountability as a relationship between and among multiple actors; and 2) political theory and the social sciences, largely focusing on work by Mark Bovens, whose framework for identifying accountability relationships has been particularly influential in contemporary scholarship on  ``algorithmic accountability'''\footnote{We discuss concerns with this phrase in Section~\ref{sec:barriers:scapegoat} (\textit{scapegoat}).}~\cite{wieringa2020account,kroll2021traceability,kacianka2021accountable}.

\subsection{Accountability in Moral Philosophy}\label{sec:rw:philosophy}

Numerous efforts in moral philosophy have sought to develop a rigorous conception of accountability. We focus on two threads in the literature, \emph{blameworthiness} and \emph{relationships between moral actors}, and correspondences between the two.

\boldedpoint{Blame}~\citet{nissenbaum1996accountability} anticipated a problem of diminishing accountability as societies become increasingly dependent on computerized systems. 
She attributed this likelihood to the emergence of barriers to accountability in computerized society, and turned to 
philosopher Joel Feinberg's work to explain how and why these barriers are prone to arise: 
Blame, defined in terms of \emph{causation} and \emph{faultiness}, is assigned to moral agents for harms they have caused due to faulty actions~\cite{feinberg1970doing,feinberg1985sua}.\footnote{Neither of these 
elements is straightforward---in fact, 
the subjects of centuries of philosophical and legal thinking. 
Faultiness, e.g., 
presumes free agency---
a concept whose metaphysical character and role in moral attribution has been the subject of 
long debate---and is a basic concept in all legal systems 
that informs judgements of legal liability (categorizing harmful actions as intentional, reckless, and negligent)~\cite{feinberg1970doing,feinberg1968collective}.} Following Feinberg, Nissenbaum conceives of actors as accountable when they step forward to answer for harms for which they are blameworthy. Her concern was that in computerized societies 
too many circumstances would arise where no one would step forward to acknowledge blame for harm, whether due to genuine puzzlement or intentional avoidance. 
Accordingly, the barriers to accountability that  
she identifies arise because the conditions of accountability are systematically obscured, due, at some times, to circumstances surrounding computerization and, at other times, to a societal breakdown in confronting willful failures. 
\textit{Many hands} obscures 
lines of causal responsibility (Section~\ref{sec:barriers:manyhands}); \textit{``bugs''} obscures the classification of errors as instances of faulty action (Section~\ref{sec:barriers:bugs}) ; \textit{scapegoating computers} obscures answerable moral actors by misleadingly or mistakenly attributing moral agency to non-moral causes (Section~\ref{sec:barriers:scapegoat}); and \textit{ownership without liability} bluntly severs accountability from blame (Section~\ref{sec:barriers:liability}) .

\boldedpoint{Relationality} 
An alternative conception of accountability expands the focus to consider responsibility in light of the relationships between moral actors. \citet{watson1996two}, for example, argues that responsibility should cover more than \textit{attributablility}, a property assigned to an actor for bringing about a given outcome \cite{talbert2019sep}. A second dimension, which he calls \textit{accountability}, situates responsibility in a \emph{relationship among actors}. For Watson, ``Holding people responsible is not just a matter of the relation of an individual to her behavior; it also involves a social setting in which we demand (require) certain conduct from one another and respond adversely to another's failures to comply with these demands"~\cite[p 229]{watson1996two}. Other work, including T.M. Scanlon's theory of responsibility, provides accounts of both \textit{being responsible} and \textit{being held responsible}, where the latter describes situations when parties violate relationship-defined norms \cite{scanlon2000we, shoemaker2011attributability}.
Accordingly, the characteristics of a harmed party might dictate whether, or what, accountability is needed. For instance, if one causes harm in self defense, there may be no moral imperative to hold them accountable.

\vspace{1.5mm}\noindent This work attempts to situate accountability in the social, political, institutional, and interpersonal relationships in which we are enmeshed. Accordingly, the relationship-defined obligations we have to one another---as spouses, citizens, employees, friends, etc.---may dictate what it is we are responsible for, as well as the types and degrees of accountability we can expect. By situating accountability not just as \textit{attributability} between action and actor, but instead within a social framework, some of what has come out of the so-called ``narrow'' notion of accountability in political theory (discussed below in Section \ref{sec:rw:narrow}) can be derived from the vantage of a more ``pure'' moral philosophy. Rather than formally pursuing this derivation here, we instead simply suggest that these notions of accountability need not be framed as alternatives to one another. Moral philosophy offers concepts through which a given relational framing---be it interpersonal, institutional, or political---can be said to be legitimate and ethically viable. Similarly, for practitioners holding a variety of organizational positions (in relation to one another), the moral responsibilities that individuals hold can shape the ethical obligations and specific forms of accountability at play.

\subsection{Accountability in Political Theory and the Social Sciences}\label{sec:rw:narrow}

The work in moral philosophy discussed above aligns with work on accountability as a property of social structures~\cite{garfinkel1984studies}, which holds it to be relational---not merely as a requirement on an accountable party to ``own up'' to blameworthy action as an obligation \textit{to} another. In the past few years, ``algorithmic accountability'' has attracted growing interest in approaches that are institutional or structural in character. The work of political scientist Mark Bovens, particularly what he has labeled, a ``narrow definition''~\cite{bovens2007analysing,bovens2014public}, has informed recent literature on accountability for ``algorithmic systems''~\cite{wieringa2020account}. Prompted by a concern that newly formed governmental structures and public authorities in the European Union lack ``appropriate accountability regimes''~\cite[p. 447]{bovens2014public}, Bovens proposed that accountability obtains between two key roles: an \emph{accountable actor} and a \emph{forum}. Under certain conditions, or in the wake of certain incidents, accountability exists when an accountable actor has an enforceable obligation to a forum to explain and justify itself---to address a forum's questions and judgments 
and possibly suffer sanctions. Bovens calls this a ``relational'' definition because it locates accountability in a social relation between those occupying one role (e.g. governmental department, a public authority, or a person acting in an official capacity) and another (e.g., a different governmental entity, oversight committee, or even an individual acting in a relevant capacity, e.g. journalist). 
We read Bovens as gesturing toward  four key parameters in any relational accountability framework for which appropriate values need to be specified: 

\textit{Who is accountable?}: \emph{Accountable actors} may include those who are not directly responsible for harm (e.g., engineers) but are designated as accountable (or liable) because of their deep pockets,  
capacities to render explanations, or positions in organizational hierarchies, such as corporate officers or government procurers of data-driven systems.

\textit{For what?}: Beyond legally-cognizable harms  
(e.g., bodily injury, property damage
), harms particularly associated with data-driven algorithmic systems include privacy violations~\cite{citron2022privacy}, automation-driven unfair discrimination~\cite{ajunwa2021auditing}, autonomy losses due to manipulation~\cite{agarwal2019deepfake,kreps2020gpt}, and any number of emergent harms associated with novel technologies and their deployment.

\textit{To whom?}: The members of the \emph{forum} may not just include those who are themselves harmed (or placed in harm's way through heightened risk). They may also include those deputized to represent and advocate on behalf of vulnerable parties, such as lawyers and public or special interest advocacy groups. Beyond direct advocates, these may include groups and individuals in oversight capacities such as journalists, elected officials, government agencies, professional societies, or the many publics which coalesce around particular matters of concern~\cite{metcalf2022relationship}.

\textit{Under which circumstances?}: This concerns the nature of the obligation---what \emph{accountable actors} may owe to the forum (to explain, be judged, and address questions and challenges). For example, \citet{moss2021assembling} describes an array of components that constitute accountability within impact assessment frameworks, noting that the specific 
obligations an actor owes to a forum depend on the norms of that relationship. 


\boldedpoint{Bringing together the moral and the relational} Proponents of Bovens's relational framework claim that it illuminates the sociopolitical stakes of transparency and explainability, 
showing why these concepts are necessary for any accountability framework for 
algorithmic societies, even though they are ultimately not sufficient to constitute accountability in and of themselves~\cite{wieringa2020account}. Moreover, by defining actors' roles and capacities in terms of the respective sociopolitical structures in which we live, Bovens's framework is \emph{not} directed at the rights and obligations we have to one another as bare moral actors. We note that bringing together Bovens's relational definition with the moral conception of accountability 
can help clarify the scope of possible values for the framework's parameters: Those who have caused or contributed to harm through faulty action are contenders for the class of \emph{accountable actors}, and those who have suffered harm (and/or their representatives) deserve a place among the members of the \emph{forum}. This point 
shows a confluence between accountability as answerability for blameworthy action, and accountability as a social arrangement. Being blameworthy for harm is (almost always) a sufficient condition for being designated an accountable actor; being harmed through blameworthy action is (almost always) a sufficient condition for being designated a member of the forum, empowered to demand 
explanations. These two conceptions do not stand against one another as alternative solutions to the same problem; they are solutions to different problems that intersect in constructive ways.

\vspace{1.5mm}\noindent Nevertheless, hard work remains to explain and justify concrete, appropriate values for these parameters, and to construct pervasive structures for accountability through context-bound  contestation~\cite{metcalf2021algorithmic}. In Section~\ref{sec:barriers} below, we demonstrate how data-driven algorithmic systems heighten the barriers to accountability by further obscuring conditions of responsibility and fault, which in turn presents challenges for instantiating the four parameters of a moral, relational accountability framework.

\section{Revisiting the Four Barriers to Accountability}\label{sec:barriers}

In a typical scenario in which software is integrated into a functional system---fully or partially displacing groups of human actors---accountability could be displaced along with human actors who 
are its bearers. The cumulative effect of such displacements is the increasing incidence of harmful outcomes for which no one answers, whether these outcomes are major or minor, immediate or long-term, or accrue to individuals or to societies. Resuscitating accountability is no simple task 
because computerization sets up particularly troublesome barriers to accountability: \textit{Many hands}~(\ref{sec:barriers:manyhands}), \textit{``Bugs''}~(\ref{sec:barriers:bugs}), \textit{The computer as scapegoat}~(\ref{sec:barriers:scapegoat}), and \textit{Ownership without liability}~(\ref{sec:barriers:liability})~\cite{nissenbaum1996accountability}. These interdependent barriers are not necessarily an essential quality of computer software. Rather, they are a consequence of how software is produced, integrated into institutions, and embedded within 
physical systems; they are a function of the wonderment and mystique that has grown around computerization, and the prevailing political economy within which the computer and information industries have thrived. In the sections that follow, we revisit the barriers 
with an eye turned toward their implications amidst the massive growth and adoption of data-driven algorithmic technologies. We provide examples of the barriers in action and defer discussion of how the barriers can be weakened to Section~\ref{sec:beyond}.

\subsection{The Problem of \textit{Many Hands}}\label{sec:barriers:manyhands}

The barrier of \textit{many hands} arises due to the large number of actors often involved in the design, development, and deployment of complex computerized systems. When such systems cause harm, it may be difficult to isolate the component(s) at its source and the agents responsible: 
``Where a mishap is the work of `many hands,' it may not be obvious who is to blame because frequently its most salient and immediate causal antecedents do not converge with its locus of decision making''~\cite[p. 29]{nissenbaum1996accountability}. Nissenbaum further analyzes the difficulty of 
\textit{many hands} by showing how it operates at four different levels: 1) software is produced in institutional, often corporate, settings in which there is no actor responsible for all development decisions; 2) within these settings, multiple, diffuse groups of engineers contribute to different segments or modules of the overall deployed system, which additionally often depends on software implemented 
by other actors (in today's landscape, this may result in licensed or freely-available open-source software); 3) individual software systems often interact with or depend on other software systems, which themselves may be unreliable or present interoperability issues; 4) hardware, not just software, often contributes to overall system function, particularly in cyber-physical systems, and it can be difficult to pinpoint if harms occur due to issues with the code, the physical machine, or the interface between the two. Any and all of these four levels of \textit{many hands} problems can operate simultaneously, further obscuring the source of blame.  

These difficulties at the heart of the \textit{many hands} problem persist, further complicated in numerous ways now that computer systems are ubiquitous rather than merely ascendant. We focus 
on how data-driven algorithmic systems complicate this barrier with novel challenges 
using two illustrative 
examples: 1) The \emph{ML pipeline}---the multi-stage process by which machine-learned models are designed, trained, evaluated, deployed, and monitored; 2) Reliance of contemporary data-driven algorithmic systems on the \emph{composability of openly-available ML toolkits and benchmarking suites}; these toolkits, often developed and maintained by large tech companies, tend to be advertised as general- or multi-purpose, and are frequently (mis)used in specific, narrow applications.

\boldedpoint{The ML pipeline} The ML pipeline is a dynamic series of steps, each of which can involve multiple groups of actors, including designers, engineers, managers, researchers, and data scientists. The pipeline typically starts with problem formulation and, in commercial settings, results in the deployment and continued monitoring of a trained model~\cite{passi2019formulation}. Problem formulation  
involves the collection, selection, or curation of a dataset, followed by the operationalization of a concrete task to learn, such as classifying loan-granting decisions or generating natural-language text. The actors responsible for formulation may hand off their work to others responsible for implementation---choosing the type of model 
and the learning procedure to use for model training. In selecting the type of model, 
these actors may custom-design their own 
architecture, or may defer to a pre-existing one, such as an off-the-shelf neural network, which has been designed by others, possibly at another company or institution. Thereafter, 
training and evaluation begin, in which a group of developers run training many times, perhaps with multiple combinations of 
model types, training procedures, and hyperparameter values. These developers compare trained models, from which they select some ``best''-performing model (or ensemble of models), where ``best'' is informed by a quantitative metric they have adopted, such as mean overall test accuracy. These stages, 
from formulation to evaluation, are often repeated dynamically: 
Until the model passes the threshold of developer-specified performance criteria, the process can cycle from re-modeling to tuning. 
If the model is deployed in practice, there is yet another set of actors who monitor the model's ongoing behavior, ensuring that its behavior 
aligns with expectations developed during training and evaluation.

Each stage of the ML pipeline involves numerous actors---in fact, potentially \emph{indefinitely} many actors if the pipeline employs third-party model architectures or ML toolkits, which we discuss below.\footnote{Participatory design further expands the set of \textit{many hands} to end-user stakeholders~\cite{sloane2020participation}, illustrating an additional manifestation of the barrier: when harms occur, it is possible to shift blame to harmed end-users who were 
involved in the ML pipeline.} Thus, in practice, if a trained model causes harms, it can be extremely challenging to tease out particular actors who should answer for them. For example, harms could originate from how actors operationalize the learning task at the start of the pipeline~\cite{cooper2021emergent}, move from high-level abstraction to concrete implementation~\cite{selbst2019abstraction}, or select hyperparameters or random seeds during model selection~\cite{forde2021disparate,sivaprasad2020hpo,cooper2021hpo}. Blame could lie with actors in any part of the pipeline, or some combination thereof whose faulty actions may have been causally responsible for harm. Bias, for example, could creep in early, from the choice of dataset, and accumulate and become magnified further downstream during model selection. In other words, the diffuse and dynamic nature of the pipeline makes locating accountability extremely challenging. This can be understood as an issue of transparency---beyond the specific the problem of model interpretability---concerning \emph{who} is responsible \emph{for what}, and \emph{how} this can be related to overarching accountability with respect to a model's ultimate use in practice~\cite{kroll2021traceability}.\footnote{This indicates why transparency in the form of model intepretability may be important, but is ultimately not sufficient, for identifying actors accountable for harms.} 

\boldedpoint{Multi-purpose toolkits} Practitioners and researchers often do not code model architectures or optimization algorithms from scratch. Just as Nissenbaum highlighted the integration of third-party software modules as the indefinite expansion of \textit{many hands}, we note here that builders of data-driven algorithmic systems  
often rely on toolkits produced by others. 
To decrease the amount of time and money spent iterating the ML pipeline, these actors depend on the investment of tech companies with vast resources and large, concentrated pools of technical talent to develop and release efficient, correct, comprehensive, and user-friendly libraries of algorithm implementations, model architectures, and benchmark datasets~\cite{abadi2015tensorflow2015,paszke2019pytorch, mattson2020mlperf}. 
Unlike more traditional modules, which only tend to contain reusable software algorithms, ML toolkits often also include large-scale, pre-trained models
. 
Large companies 
train and release such models, like 
BERT~\cite{devlin2019bert}, which smaller companies and individuals 
can use out-of-the-box or fine-tune for particular use cases. Since these pre-trained models are often intended for downstream use by users different from their developers, they are designed 
for a multiplicity of applications (i.e., to be general-purpose). However, users employ pre-trained models in specific 
domains; there is a gap between general design goals and specific deployment intentions, which has been shown can bring about bias-related harms. Determining blame for these types of harms is far from simple. For example, if intended use is under-specified, blame could lie at least partially with the pre-trained model's creator. Compounding this problem is the fact that ML presents a recursive turn in the \textit{many hands} problem Nissenbaum highlighted, in that many ML systems incorporate pre-trained components that are, themselves, the product of \textit{many hands}. Nevertheless, tracing such harms presents an addressable technical challenge, not an insurmountable epistemological barrier.

\vspace{2mm}\noindent\textbf{In relation to a moral, relational accountability framework, this barrier obscures \ldots}\vspace{1mm}

\textit{Who is accountable}: 
\textit{Many hands} is central to identifying an accountable actor within Bovens's framework~\cite{bovens2007analysing}. This problem 
has long characterized challenges in holding corporate actors, institutions, and organizations accountable, and while it certainly constituted a barrier to accountability 
in 1996~\cite{nissenbaum1996accountability}, it has only become more difficult to understand who is accountable in an algorithmic society. 
Code reuse---
taken as a virtue in software development---
has now been extended to model reuse, in turn generating a host of problems for  
equity and reliability by making it difficult to identify all the actors who contributed to components of an ML pipeline. Knowing who is responsible for these components as they are repurposed, as well as who ought to be responsible for incorporating those components into a downstream system, becomes prohibitively difficult for a forum to ascertain on its own, let alone for it to demand any explanations or changes in actors' behavior. 

\textit{For what}: The problem of \textit{many hands} extends the above question to determining \textit{what} an actor might be accountable for in relation to 
harms, in that it is hard to isolate which part of an ML pipeline actually contributes to an error or harm. Repurposed models may introduce dataset imbalances and proxies for protected categories 
without adequate scrutiny (or even the opportunity for scrutiny) by those assembling downstream components of a system. This raises  questions of appropriate use, wherein it is difficult to tease apart the responsibility of those who produced a component to adequately stipulate the limits of its appropriate use and the responsibility of those who use that component to ensure it is appropriate for the uses to which they are putting it. 

\textit{To whom}: 
\textit{Many hands} is primarily a barrier to knowing \emph{who} is accountable, but it is also a barrier to knowing \emph{to whom} those accountable actors are accountable where, for example, a differential error rate may exist for some population $\mathcal{P}$, but a specific harm occurs for an individual $p \in \mathcal{P}$. In such a case, it is difficult to determine whether accountability ought to be rendered to $\mathcal{P}$, because of the heightened risk of harm to which the entire population has been exposed, or only to $p$, who suffered harm because of their membership in $\mathcal{P}$. This is a \emph{many hands} problem because of the difficulty in knowing where within the ML pipeline risk was produced for the group, e.g. through training or dataset imbalances, and where it was produced for individuals, e.g. through implementation choices. 

\textit{Under which circumstances}: The problem of 
\textit{Many hands} presents a barrier 
even when standards of care exist, as it is difficult for actors to know precisely whom they should exercise that care toward (see ``to whom'' above). Standards of care, which are grounded in normative assumptions about appropriate component (re)use, 
are less straightforward to develop where \textit{many hands} are involved, as social practices which link actors together are obscured throughout the ML pipeline. 


\subsection{\textit{``Bugs''}}\label{sec:barriers:bugs}

Nissenbaum uses the term \textit{``bug''} to cover a variety of issues common to software, including ``modeling, design, and coding errors.'' \textit{``Bugs''} are said to be ``inevitable,'' ``pervasive,'' and ``endemic to programming,'' ``natural hazards of any substantial system''~\cite[p. 32]{nissenbaum1996accountability}. Even with software debuggers and verification tools that can assure correctness, \textit{``bugs''} emerge and cause unpredictable behavior when software systems are deployed and integrated with each other in the real world~\cite{smith1985limits, mackenzie2001proof}. The rhetorical power of \textit{``bugs''} is that they are predictable in their unpredictability; they serve as a barrier to accountability because they cannot be helped (except in obvious cases), and therefore are often treated as an accepted ``consequence of a glorious technology for which we hold no one accountable''~\cite[p. 34]{nissenbaum1996accountability}. 
What we consider to be the ``inevitable'' can change over time as technology evolves, with certain types of \textit{``bugs''} spilling over into the avoidable. For example, evolving norms and new debugging tools can rebrand the ``inevitable'' to be sloppy or negligent implementation, at which point programmers can be held to account for such errors. Similarly, the advent of data-driven algorithmic systems has 
indicated that this malleability also extends in the other direction: New technological capabilities can both contract and expand what we consider ``inevitable'' \textit{``buggy''} behavior. That is, while these systems contain \textit{``bugs''} of the ``modeling, design, and coding'' varieties that Nissenbaum describes for rule-based programs, the statistical nature of data-driven systems presents additional types of harm-inducing errors, which may present an additional barrier to accountability.\footnote{Of course, statistical software is not new to ML; however, the proliferation of data-driven algorithmic systems has clarified the prevalence of such errors.} 
Where misclassifications, statistical error, and nondeterministic outputs cause harm---and are presented as inevitable and unavoidable --- may impede the attribution of blame.

In 1996, it may have been evident that labeling certain errors as \textit{``bugs''} was a mere ploy to dodge blame. Today, certain types of errors are more plausibly asserted to be an inherent part of ML, attributable to its statistical nature. Misclassification, statistical error, and nondeterminism seem to turn the notion of \textit{``bug''} on its head: Indeed, many experts would as readily call these \textit{features} of machine learning, 
not \textit{``bugs''}.\footnote{We return to this in Section~\ref{sec:barriers:scapegoat} (\textit{scapegoat}) 
and is 
why we leave \textit{``bugs''} in quotes.} Nevertheless, regardless of where one attempts to draw the line, these errors share common elements with the \textit{``bugs''} Nissenbaum describes---namely, they undermine our ability to reason, conclusively, about causality and fault. Insofar as they are accepted as an ``inevitable,'' ``pervasive,'' and  ``consequence of a glorious technology,” they constitute a barrier to accountability~\cite{nissenbaum1996accountability}. Below, we illustrate this point with concrete instances of \textit{``bugs''} deemed unavoidable in data-driven algorithmic systems.

\boldedpoint{Faulty modeling premises} As discussed in Section~\ref{sec:barriers:manyhands}, data\--driven algorithmic systems require significant modeling decisions prior to implementation. For example, choosing a model to learn necessarily involves abstraction and can have significant ramifications~\cite{selbst2019abstraction, passi2019formulation}. Assumptions during this stage of the ML pipeline can bias the resulting computational solution space in a particular direction~\cite{friedman1996bias}, for example, assuming a linear model is sufficient to capture patterns in data precludes the possibility of modeling non-linearities. When such biases involve over-simplified or faulty reasoning, they can result in model mis-specification and the introduction of ``modeling error bugs.'' Such mis-specifications may include the assumption that values like fairness and accuracy are correctly modeled as a trade-off to be optimized~\cite{cooper2021emergent},  and that physical characteristics can serve as legitimate classification signals for identifying criminals~\cite{wu2016automated} or inferring sexual orientation~\cite{stark2021pseudo, wang2018gay}. More generally, a common modeling error may arise from assuming, in the first place, that a problem is amenable to classification---that it is possible to divide data examples into separable categories~\cite{sun2019gender, sloane2021silicon}. Even if it is possible to train mis-specified models like these to behave ``accurately'' (i.e., to return better-than-chance results after learning these tasks), conclusions drawn from false premises will be unsound~\cite{cooper2021emergent}. If modeling assumptions are unclear or elided, an actor may evade accountability by blaming inexplicable, unavoidable \textit{``bugs''} endemic to computer software instead of taking responsibility for otherwise opaque errors.

\boldedpoint{Individual errors} Even if one’s premises are not faulty, the ML pipeline can still produce models that cause harm. Trained ML models exhibit errors that can harm individuals if their effects, for example, violate privacy or cause manipulation~\cite{feldman2015disaparte, lovering2021predicting, nadeem2021stereoset, kreps2020gpt}. ML has several techniques to quantify and minimize error~\cite{botchkarev2019metrics,hardt2016equality}, and yet 
even the most robust, well-trained models report imperfect accuracy. In fact, a model that achieves $100\%$ accuracy is usually considered suspect, likely over-fit to the training data and to exhibit poor performance when presented with new examples~\cite{srivastava2014dropout, hastie2009statistical, recht2019imagenet}. Therefore, when individual errors occur, they can be treated as inevitable, just like the \textit{``bugs''} Nissenbaum describes, displacing responsibility for the harms such errors cause affected individuals.

\boldedpoint{Bad model performance} Unexpectedly bad overall model performance can likewise be excused as a \textit{``bug,''} 
rather than a blameworthy error. Consider a hypothetical example of a (well-formulated) computer vision system used to detect skin cancer, whose training and evaluation indicate will have an 
accuracy rate of 94\%. Once deployed, if the model coheres with (or even out-performs) its promised performance, then developers can claim that any mis-classifications were 
expected.\footnote{Individual errors 
can pose additional challenges for accountability: 
The model may still overall exhibit an expected degree of error (i.e., be within a margin of error), for which it is possible to scapegoat 
the 
statistical nature of ML (Section~\ref{sec:barriers:scapegoat}).
} Since expected accuracy is a probabilistic claim about what is likely to occur, deviations from expectation can and do occur. When monitoring a deployed model, over time, if this deviation yields a substantial decrease in expected 
accuracy, developers may dodge accountability by ascribing the failure to the amorphous category of \textit{``bug''}, instead of admitting that it resulted from human negligence, poor generalization, distribution shift, or other faulty behavior.

\vspace{2mm}\noindent\textbf{In relation to a moral, relational accountability framework, this barrier obscures \ldots}\vspace{1mm}

\textit{Who is accountable}: Accountability for \textit{``bugs,''} even within the expanded definition of \textit{``bugs''} provided above, emerges from specific regulatory regimes, corporate compliance practices, and contracting relationships. Civil law has a crucial role in determining the relationship between forums and responsibilized actors, which is often inflected by those who have the capacity to intervene or have benefitted from a particular action. \textit{``Bugs''} present a particular challenge to determining who is accountable when they are seen as endemic to ML, or as produced by non-determinism inherent to the domain in which an algorithmic system is deployed (a challenge shared by the \emph{scapegoating} barrier).

\textit{For what}: \textit{``Bugs''} remain a barrier 
because of the difficulty they pose to actors and forums trying to specify whether individual errors, bad model performance, faulty assumptions, or other 
mistakes contributed to a harm.

\textit{To whom}: \textit{``Bugs''} may affect an entire class of individuals, a community, or all of society, but evidence of harm may only accrue at the level of a specific individual, presenting a barrier for actors and forums interested in knowing to whom accountability ought to be rendered.

\textit{Under which circumstances}: Algorithmic systems inevitably rely on some degree of abstraction and make specific assumptions about the underlying nature of the phenomena they model~\cite{selbst2019abstraction,cooper2021emergent}. Under circumstances of imperfect information about every possible aspect of a data-driven algorithmic system (which is most of the circumstances outside the lab), \textit{``bugs''} of the character described above may exist and contribute to this barrier to accountability.

\subsection{\textit{The Computer as Scapegoat}}\label{sec:barriers:scapegoat}

Blaming a computer may pose a barrier to accountability, because ``having found one explanation for an error or injury, the further role and responsibility of human agents tend to be underestimated" \cite[p. 34]{nissenbaum1996accountability}\cite{elish2019crumple}. To explain why people could plausibly blame computers for  wrongdoing, Nissenbaum cites the role computers may play in ``tasks previously performed by humans in positions of responsibility;'' whereas before the human would be indicated as the blameworthy party, the computer has now taken up that role. And yet, even as computer systems have become immediate causal antecedents to an increasing number of harms, they 
lack moral agency and thus cannot be the bearers of moral blame~\cite{nissenbaum1996accountability}. In this section, we discuss how \textit{scapegoating the computer} has become even more complicated in the landscape of ubiquitous data-driven algorithmic systems. 
In the examples below, the 
system is made to bear the sins of the responsible party, the individual or the institution that has agency and is capable of carrying moral blame.

\boldedpoint{Moral agency}
As data-driven algorithmic systems have become pervasive in life-critical contexts, there has been a corresponding tendency to anthropomorphize and view technological processes as akin to human cognition~\cite{turkle2005second, pradhan2019phantom, birhane2020robot}.  These systems are described by their developers and commentators as intelligent, implying that they have agency as autonomous actors and thus 
rhetorically positioning them as 
blameworthy for error. 
However, directing blame toward data-driven algorithmic systems effectively imbues them with moral agency, ascribing them the ability to act intentionally~\cite{schlosser2019sep}. 
Nissenbaum likens blaming a computer to blaming a bullet in a shooting: While the bullet can be said to play an active, causal role, it cannot be said to have been intentional in its behavior. In the same vein, a data-driven algorithmic system may play a central role in life-critical decisions, and may even be said to \textit{make a choice} in a particular task, but a choice lacking deliberate intention, a precondition for moral agency~\cite{schlosser2019sep}.\footnote{This is consistent with scholarship in 
legal theory concerning AI, algorithms, agency, and personhood~\cite{veliz2021moral, birhane2020robot, himma2009artificial, bryson2017and, calo2015robotics, balkin2015robotics, lemley2019remedies}.} 

\boldedpoint{``Accountable algorithms''} This popular banner-phrase 
makes \textit{algorithms} the subject of accountability~\cite{kroll2017accountable}, 
even though algorithms are not bearers of moral agency and, by extension, moral responsibility. It places 
responsibility on 
technology, not its developers, owners or operators, and it reduces accountability to a piecemeal, procedural quality that can be inferred from technology, rather than a normative concept that has to do with the moral obligations that people have toward one another. The phrase further occludes proper attribution of accountability by fixating attention on algorithms rather than on systems that are deployed in practice, within and through which algorithms function~\cite{cooper2021eaamo}. 
When, for example, studies of fairness in AI/ML-assisted judicial bail decisions fixate on respective algorithms, they fail to capture key inequities that are systemic in complex sociotechnical systems, of which AI/ML techniques are just one part~\cite{barabas2020studying}.

\boldedpoint{Mathematical guarantees} Directing blame away from people and corporations can be either strategic or inadvertent. In some cases, a group of harmed individuals does not know whom to blame (\textit{many hands}) and settles on blaming the system. In others, scapegoating the system can be a way by which a moral actor dodges and dissipates public ire, for example, in the now-canonical example of Northpointe exhibiting bias in its risk-assessment tool~\cite{angwin2016machine}. Rather than attributing this bias to a mistake or \textit{``bug,''} Northpointe blamed the fundamental incompatibility of different algorithmic operationalizations of fairness as the source of the problem (and pointed to a specific measure, for which bias was not detectable, as evidence of blamelessness). Reliance on mathematical guarantees can reinforce barriers to accountability and divert attention away from its appropriate subjects. One can see this when a given system has a theoretically-guaranteed (and empirically-verified) upper bound on its error. If the system behaves within its guaranteed margin of error, it becomes possible to treat that margin as an immutable attribute of the system (rather than, more appropriately, the result of human-made decisions), and to scapegoat the system for any particular errors that fall within this margin. 

Let us consider the same case we discussed for the problem of individual errors in \textit{``Bugs:''} The engineers show that a system is 94\% accurate for tumor detection, and validate that this is in fact the case in practice. Above, we talked about this example in terms of individual errors, for which responsibility for harm could be excused due to \textit{``buggy''} behavior. Rather than analyzing behavior at this level of individual decisions, one can also examine the behavior \emph{of the model overall}. If the frequency of mis-classifications is within the model's guaranteed error rate, the engineers could attempt to excuse all resulting harms by gesturing to the fact that the model is performing \emph{exactly as expected}. In short, satisfying mathematical guarantees can serve as a \textit{scapegoat} because pointing to mathematical claims satisfied at the model-level can serve to obscure the need to account for harms that occur at the individual-decision level.\footnote{One could  
see-saw back-and-forth between \textit{``bug''} and \textit{scapegoat} to evade accountability. If satisfying guarantees at the overall model-level is 
rejected as a rationale for an individual harm, one could claim there is a \textit{``bug;''} if calling an individual decision \textit{``buggy''} is rejected, and the model is classifying within its expected error, one could then displace blame by arguing that the model is behaving according to its specification.}

\boldedpoint{Non-determinism} When data-driven algorithmic systems err, their errors can be attributed to the stochastic, non-deterministic components of either the system itself or the phenomena the system is modeling. In particular, 
systems that involve ML 
involve randomization, 
for example, by shuffling the order in which training data examples are presented to an algorithm. While such features of ML algorithms may seem like technical minutiae, in fact, they introduce stochasticity into the outputs of machine-learned models: Training the same model architecture on the same dataset with the same algorithm---but changing the order in which the training data are supplied to the algorithm---can yield models that behave differently in practice. For example, as ~\citet{forde2021disparate} shows, changing the order that the data examples are presented to train a tumor-detection model can lead to surprisingly variable performance. The relationship between training-data-ordering and resulting variance in model performance is under-explored in the technical literature. Thus, such differences in model performance are often attributed to an inherent stochasticity in ML. The randomization used in ML systems---randomization on which these systems depend---becomes a \textit{scapegoat} for the harms it may cause, such as missed tumor detection. In attributing the harms to mathematical chance, attention is drawn away from appropriate accountable agents.

\vspace{2mm}\noindent\textbf{In relation to a moral, relational accountability framework, this barrier obscures \ldots}\vspace{1mm}


\textit{Who is accountable}: Similar actors are accountable as those described in \textit{``Bugs,''} although the barrier 
presented by \textit{scapegoating} is embedded in its implicit suggestion that entities are accountable, rather than those who are the responsible actors (see the nebulousness of \textit{many hands}), or that no responsible actor can be found because a harm occurred through randomness or chance.

\textit{For what}: \textit{Scapegoating} produces barriers to understanding the \emph{for what} of accountability in identical ways as described above in \textit{``Bugs,''}. It also contributes an additional difficulty when mathematical guarantees are offered that allow for some minimal degree of undesirable behavior in a system, or the system is characterized as non-deterministic in ways that would indemnify otherwise responsibilized actors from accountability for outcomes stemming from such undesirable behaviors.

\textit{To whom} and \textit{under which circumstances}: Same as in Section~\ref{sec:barriers:bugs}.



\subsection{Ownership without Liability}\label{sec:barriers:liability}

\citet{nissenbaum1996accountability} highlights a dual trend in the computer industry: 1) strengthening property rights and 2) avoiding liability. Behavioral trends that informed these assertions have persisted in the decades since, with lively public debates over the fit of traditional forms of intellectual property (i.e., copyrights, patents, and trade secrets) to digital products such as software, data, databases, and algorithms~\cite{cofone2019secrecy, fromer2019machines}, and subsequent expensive legal struggles among industry titans~\cite{harvard2021oracle}. Similarly, we have seen explicit denials of liability expressed in shrink-wrap licenses, carried over into so-called ``click-wrap'' licenses, and Terms of Service disclaimers accompanying websites, web-based services, mobile apps, Internet of Things devices, content moderation decisions, and the like~\cite{kosseff2022sec230, citron2022privacy, levy2014intimate, tereszkiewicz2018digital}.

Before addressing how we see these trends carry forward in the contemporary landscape, we need to qualify our observations. Property and liability are weighty legal concepts with long histories and rich meanings. Narrowing our view to digital technologies, even before \citet{nissenbaum1996accountability}, a robust literature had grown over questions of ownership---questions that have persisted through numerous landmark court cases. Liability, too, is a core legal concept that is increasingly an issue in relation to the products and services of digital industries. It lies outside the scope of this paper to attempt meaningful insights into these concepts as they manifest in scholarship, law, and the courts. However, it is useful to observe broad patterns and anticipate the likely actions of stakeholders. For a start it is not difficult to see how the trends toward strong ownership and weak liability reinforce barriers to accountability, and also to understand why industry incumbents might support them: Liability is costly and strong property rights enrich rights holders and empower them against competitors. Four lines of advocacy on behalf of industry interests are noted below, supplementary to those discussed in~\citet{nissenbaum1996accountability}:
\begin{itemize}[topsep=0pt, leftmargin=.4cm]
\item Third-party providers of data-driven algorithmic systems refuse to expose their systems to scrutiny by independent auditors on grounds of trade secrets~\cite{cofone2019secrecy, fromer2019machines}. As long as experts maintain that transparency is necessary to evaluate the ML pipeline and AI development, strong property rights that block scrutiny are barriers to accountability.
\item Manufacturers and owners of cyber-physical systems, such as robots, Internet of Things devices, drones, and autonomous vehicles, evade liability for harms by shifting blame to environmental factors or humans-in-the-loop~\cite{lemley2019remedies}. In this respect, the barrier of \textit{ownership without liability} for data-driven algorithmic systems suggests a twist on the problem of \textit{scapegoating} (Section~\ref{sec:barriers:scapegoat}): treating ``the human user as scapegoat''---claiming the user has mis-used an AI- or ML-enabled system in order to obscure responsibility for unclear, under-specified, or deliberately misleading user interfaces or expected use, as has happened with Tesla and accidents concerning its (so-called) ``AutoPilot'' autonomous driving feature~\cite{boudette2021tesla}.
\item Almost without question the computer industry, having metamorphosed into the data industry, has assumed ownership over data passing through its servers~\cite{ftc2014databrokers,okidebe2022data, lambdan2019databrokers}. We still do not have clear rules of liability for industry actors when their servers, holding unimaginable quantities of data, are breached~\cite{sharkey2016can}. Nor do we have sufficient insight into the completeness, quality, or validity of data, or the means to hold anyone liable for its misuse.
\item Technology companies hold unprecedented sway over regulation. Twenty-five years ago, the software industry was already a force to be reckoned with and successfully persuaded Congress that imposing legal constraints would stifle innovation---that societal well-being depended on a nascent industry that could not flourish under excessive regulatory and legal burden. 
\end{itemize}

\vspace{2mm}\noindent\textbf{In relation to a moral, relational accountability framework, this barrier obscures \ldots}\vspace{1mm}

\textit{Who is accountable}: Having already enumerated above the many difficulties these barriers pose for tracing relationships of accountability, they generally pertain to the problem of \textit{ownership without liability}, as well. Additionally, questions of how liability is adjudicated in practice may obscure who is liable, what kind of liability they hold, or what they are liable for, while leaving intact the ways in which the benefits of data-driven algorithmic systems accrue to their developers, designers, and operators.

\textit{For what}: 
\textit{Ownership without liability} affects the very contours of what an actor can be found liable for. However, this does not absolve that actor of their moral responsibility or obviate the need for them to be held accountable for the consequences of their actions, the systems they oversee, or from which they benefit.

\textit{To whom}: \textit{Ownership without liability} is a barrier to accountability for those who may stand as plaintiffs in civil cases and representatives of those affected.

\textit{Under which circumstances}: \textit{Ownership without liability} is a barrier 
where those who suffer a harm lack standing in a court of law. This may be because a harm is not cognizable to courts (see, e.g.,~\citet{metcalf2022relationship}), the harmed party does not constitute a certifiable class, or the nature of the harm is obscured through the ways harms are foisted onto \textit{scapegoats} or dismissed as \textit{``bugs''}.  


\section{Weakening the Barriers}\label{sec:beyond}

\citet{nissenbaum1996accountability} warned of a waning culture of accountability---harms befalling individuals, groups, even societies, were being cast merely as sufferers' bad luck. In the previous section, we revisited the four barriers in light of data-driven algorithmic systems and found that the framework still provides a useful lens through which to locate sources contributing to the dissipation of accountability. Weakening the barriers would clear the way for more sound attribution of blame, in turn setting up a stronger societal expectation for blameworthy parties to step forward and take account. But we have also argued that accountability in algorithmic societies involves more: Stepping forward is a necessary component of accountability, but it is insufficient (Section~\ref{sec:framing}). Because the barriers we have described may not all be weakened, even with a firm resolve to identify blameworthy parties, we need more than astute attention on a case-by-case basis. To build a lasting culture of accountability, a necessary supplement involves establishing persistent institutional frameworks for identifying accountable parties (i.e., individuals, groups, or organizations) and for calling them to answer. Simultaneously, such frameworks should invest others with the powers to call these parties to account.

Any technical interventions that the research community has already developed---notably, those that we have emphasized concerning transparency, audits, and robustness---would need to be folded into such a framework, and their use justified in these moral and relational terms. For example, any technical definition of transparency is unlikely to satisfy the needs of all those who comprise a forum and who may hold variable or inconsistent ideas about what it might mean for a model to be ``interpretable.'' Technical assertions of robustness say what expectations are, but leave unanswered the question of the conditions under which deviations from expectations ought to be expected or remedied.\footnote{Moreover, if assumptions underlying such assertions are voided when moving from theory to deployment, robustness estimates can degrade in practice.} Relational treatments of these issues, it would seem, require that the obligation be tuned to the various needs of all members of the forum.

\boldedpoint{Taking each barrier in turn} A moral and relational accountability framework opens the aperture to addressing \textit{many hands} 
(Section~\ref{sec:barriers:manyhands}). In principle, many, if not all, of the \textit{many hands} could be designated as accountable actors. Deliberate consideration of the \textit{many hands} problem is clearly called for by those who develop licensing agreements relying on normative assumptions about 
appropriate use and reuse within the ML pipeline, and in articulating engineering best practices empirically against theoretical assumptions of robustness. This 
includes dataset creators, model developers, decision and control systems designers, vendors, and operators of these systems. Developing rigorous standards of care could help mitigate the problems of inappropriate use of pre-trained models and unclear measures of quality control at different stages of the ML pipeline. For example, robust auditing mechanisms at each stage, rather than approaching audit as an end-to-end concern~\cite{raji2020closing}, or worse, as a purely \textit{post hoc} endeavor, could help clarify the relationship between stage-specific issues and resulting harms.

Addressing various harms, depending on how they are contextualized, can implicate either the barrier of \textit{``bugs''} or \textit{scapegoating the computer} (Sections~\ref{sec:barriers:bugs} \&~\ref{sec:barriers:scapegoat}). For example, we note that the computer science community could have either treated algorithmic harms due to unfair discrimination as a \textit{``bug''} or blamed them on intrinsic aspects of AI/ML---and yet, it did not. Instead, unfairness has more often been ascribed to biased or imbalanced training data~\cite{kallus2018residual, fish2016fair}---data that exhibits historical biases that are arguably ``pervasive'' and ``unavoidable.'' This community could have pursued some ``tolerable'' degree of unfavorable outcomes in the real world (ideally, in consultation with those adversely impacted), and developed ways of ensuring models met that more ``tolerable'' specification, under specific conditions.  
This, notably, would still have allowed developers to evade accountability by \textit{scapegoating} inherent properties of the model as instead deserving of blame. 

However, instead of treating unfairness as an aspect of accountability, much technical work on algorithmic fairness has 
attempted to address unfairness harms by developing training algorithms that are robust to biased input data. The field of algorithmic fairness therefore serves as an example that challenges the narrative of the invulnerability of the barriers. The technical community and its interlocutors have demanded more from ML modelers concerning the treatment of unfair discrimination. 
The community has set expectations concerning the necessity of interventions to root out and correct for unfairness, thereby weakening the barriers of \textit{scapegoating} or being attributed to \textit{``bugs''}. This example could, and we believe should, encourage similar treatment of other issues like robustness and its relationship to privacy violations, or adversarial ML and its relationship to manipulation. 

Lastly, \textit{being liable} is related but not identical to being accountable (Section~\ref{sec:barriers:liability}). The latter is applied to blameworthy parties who step forward to answer, the former to parties who step forward to compensate victims of harm. Often liability is assigned to those who are found to be blameworthy. If lines of accountability are blurred, for example, as a consequence of the barriers we have discussed, harms due to AI/ML and other data-driven algorithmic systems will be viewed as unfortunate accidents; the cost of ``bad luck" will settle on victims. Instead, legal systems have developed approaches, such as strict liability, to compensate victims harmed in certain types of incidents even without a demonstration of faulty behavior. Strict liability assigned to actors who are best positioned to prevent harm is sound policy as it motivates these actors to take extraordinary care with their products. Barriers such as \textit{many hands} make the attribution of blame difficult. Strict liability for a range of harms that are produced by \textit{many hands} would shift the ``bad luck" from victims to those best positioned to mitigate and prevent such harms.

\vspace{1.5mm}\noindent Eroding the barriers of accountability is a key societal challenge requiring multiple forms of expertise and, with respect to 
ML especially, the use of these tools needs to be justified. Just as mature political governance requires durable institutions and formal attributions of rights and duties, we have similar needs for the governance of producers, purveyors, and operators of data-driven algorithmic systems. That is, as we have contended throughout this paper, accountability is moral \emph{and} relational. It depends on social, legal, and political structures that provide legitimacy for the checks actors and forums place on each others' behavior; it depends on the way those checks are internalized as professional, personal, legal, and ethical duties that motivate actors' personal responsibility. Multi- and inter-disciplinary research on 
accountability, fairness, and transparency---given its potential to bring together an array of expertise focused on themes of 
equity and justice---is uniquely positioned to help develop a moral, relational accountability framework. Such structures provide legitimacy, as well as the professional codes and standards of care, disciplinary norms, and personal mores that tie moral and relational forms of accountability together. The future work of creating these structures, as noted earlier, is no small undertaking, it lies in the sociopolitical contestations, the hard, deliberative work of living within a pluralistic society, by the many constituencies implicated in any particular computational system.

\section{Conclusion}\label{sec:conclusion}
In this paper we revisited Nissenbaum's ``four barriers'' to accountability, with attention to the contemporary moment in which data-driven algorithmic systems have become ubiquitous in consequential decision-making contexts. We have drawn on conceptual framing from Nissenbaum's use of the concept of \emph{blameworthiness} and how it can be aligned with, rather than cast in opposition to, 
Bovens's work on accountability as a \emph{relational property of social structures}~\cite{bovens2007analysing,bovens2014public}. We have demonstrated how data-driven algorithmic systems heighten the barriers to accountability with regard to determining the conditions of blame, and have looked ahead to how one might endeavor to weaken the barriers. In particular, 
we have put forward the conditions necessary to satisfy a moral and relational accountability framework, discussed how the development of such a framework would weaken the barriers, and argued that an interdisciplinary and multidisciplinary research community is uniquely positioned to construct such a framework and to develop lines of inquiry to erode the barriers to accountability. 

Given our tender historical moment, addressing why these or those parties belong in the forum or in the set of accountable actors, why those obligations are justified, and, of course, evaluating the numerous permutations the relational nature of the approach demands is the provenance of future work. No easy formulations make sense until we have developed a rigorous approach to justification. In our view, this calls for expertise in relevant technologies, moral philosophy, the prevailing political economy of data and computing industries, organizational sociology, current political and regulatory contexts, domain area expertise, and more. It is not that all these are needed all the time; but any of them may be called in to develop linkages between proposed values and social welfare.


\begin{acks}
The authors would like to thank the Digital Life Initiative at Cornell Tech for its generous support. A. Feder Cooper is additionally supported by the Artificial Intelligence Policy and Practice initiative at Cornell University and the John D. and Catherine T. MacArthur Foundation. Benjamin Laufer is additionally supported by NSF Grant CNS-1704527. Helen Nissenbaum is additionally supported by NSF Grants CNS-1704527 and CNS-1801307, and ICSI Grant H98230-18-D-006.
\end{acks}

\bibliographystyle{ACM-Reference-Format}
\bibliography{bibliography}

\end{document}